\documentclass{eas}
\usepackage{graphicx}
\newcommand{\rdeV}{\mbox{$r^{1/4}$}}
\newcommand{\rSer}{\mbox{$r^{1/n}$}}

\newcommand{\Cindex}{\mbox{$C_{r_e}(\alpha)$}}
\newcommand{\apj}{ApJ}
\newcommand{\apjl}{ApJL}

\newcommand{\aj}{AJ}
\newcommand{\mnras}{MNRAS}
\newcommand{\aap}{A\&A}
\newcommand{\aaps}{A\&AS}
\newcommand{\apss}{Ap\&SS}

\TitreGlobal{JENAM 2002 Galaxy Dynamics Workshop, Porto, Portugal, 2002
September 3-7}

\begin{document}

\title{Spheroid and disk signatures in galaxy bulges}
\author{Marc Balcells}\address{Instituto de Astrof\'\i sica de Canarias, 
38200 La Laguna, Tenerife, Spain (balcells@ll.iac.es)}
%
%
\begin{abstract}
Recent progress on the structure and dynamics of bulges is reviewed.
Those aspects that link galaxy bulges either to oblate spheroids akin to
elliptical galaxies or to rapidly-rotating, flattened systems more nearly
resembling the products of disk internal transformations, are highlighted.  The
analysis of surface brightness profiles derived from HST data is reviewed to
show that unresolved nuclear components detected by HST have biased the
determination of surface brightness profiles obtained from the ground; 
\rdeV\ profiles are virtually nonexistent in galaxy
bulges.  Predictions from accretion N-body models on the shape  of surface
brightness profiles are discussed.    The position of bulges on the Fundamental
Plane (FP) of elliptical galaxies is examined to infer clues on bulge
population ages and bulge dynamical structure. Kinematic diagnostics on the
internal dynamics of bulges are examined.   Finally, a new approach to the
kinematic analysis of galaxies, based on the use of synthetic spectra of single
stellar population models instead of the standard stellar templates, is
presented.
\end{abstract}
\maketitle
\section{Introduction}
\label{MB:Introduction}

In mature disk galaxies of early to intermediate Hubble type, the central
regions {\sl bulge} above the disk: in edge-on views, the isophotes defining
the disk bend up when approaching the galaxy center.  Such central {\sl bulges}
have other distinct properties; foremost, they have a  smooth appearance in
edge-on views, implying little dust and no current star formation activity
outside of the plane of the disk.  This motivates the notion of bulges being a
separate galaxian component, with structure, dynamics, populations and
formation history distinct from those of the disk.  The prominence of the
central bulge is an important discriminant of the Hubble sequence, the other
being the morphology of the spiral arms (Sandage \& Bedke 1994; see Gilmore
1999).  

The high luminosity densities of bulges make them accessible to observation
even at high redshift.  On the other hand, their small sizes relative to those
of the disk (ratio of scale sizes $r_e/h \sim 0.1$; Courteau et al.\ 1996;
Graham 2001, hereafter G01; MacArthur et al.\ 2003), make the study of the
inner structure of bulges difficult; in small bulges, $r_e$ is comparable to
the scale height of the disk.  

Work in the eighties led to the view that galaxy bulges are in most ways
identical to elliptical galaxies that acquired a stellar disk around them. 
These similarities supported bulge formation models based on the monolithic
collapse paradigm, or on mergers as explored by, e.g., Kauffmann (1996).  
During the past decade, observational results and theoretical modeling have
provided new clues to re-address the basic questions on the nature of bulges:
How similar are bulges to elliptical galaxies? How old are bulges? Is bulge
formation prior to or after disk formation?  Do bulges grow such that galaxies
evolve along the Hubble sequence? 

Recent extensive reviews on bulges are found in Wyse et al.\ (1997) and Carollo
et al.\ (1999).  In this paper bulge properties are reviewed to highlight
evidence for the spheroid vs.\ disky nature of bulges.  Since the Kormendy
(1993) review, this field has grown, mostly due to HST imaging work and to new
kinematic studies.  

\section{Definitions of bulges}
\label{MB:Definition}

In \S\,\ref{MB:Introduction} we gave the original definition of a bulge: a
smooth light distribution that protrudes from the central
part of the disk in highly-inclined galaxies.   This definition emphasizes
the old age of the bulge stars, the lack of both star formation and patchy dust. 
To implement this definition in practice, we look for a central region with
lower ellipticity than that of the disk.  This isophotal criterion was used by
Kent (1986) and by Andredakis et al.\ (1995, hereafter APB95).  

Component analysis of galaxy structure (Kormendy 1977) introduced a powerful,
easy-to-apply methodology to characterize bulges via fitting analytic surface
brightness profile models such as the \rdeV\ model.   This led to a {\sl
surface-brightness definition} of bulges as the extra central light above the
inward extrapolation of the constant-scale length exponential disk.  The STScI
workshop on bulge formation (Carollo, Ferguson \& Wyse 1999) took this as the
operational definition of a bulge.   Widely used today, this definition might
not be ideal, as it leads to assigning to the bulge any inner brightening of
the disk.  

The co-existence of two definitions (isophote-based and surface
brightness-based) for bulges is less than ideal.  While the isophotal definition
associates bulges with spheroidal structures, and suggests formation models
similar to those of other galaxian spheroids such as elliptical galaxies,  the
surface-brightness criterion allows one to find {\sl bulges} in late-type dwarf
galaxies, as long as the surface brightness profile shows a central excess with
respect to an exponential law.  Most likely, the latter bulges do not share the
same formation mechanism as the large bulges of early-type galaxies.  

\section{Early results}
\label{MB:Eighties}

At the end of the eighties, bulges were understood to follow the \rdeV\ surface
brightness profile (Simien \& de Vaucouleurs 1986).  Early spectroscopy
(Kormendy \& Illingworth 1982) placed large bulges along the oblate rotator
line in the rotational support $V/\sigma_0$ vs $\epsilon$ diagram (some bulges
in barred galaxies were found above the oblate-rotator line; Kormendy 1982). 
The stellar populations of bulges were understood as having metallicities
higher than solar, after the pioneering work by Whitford (1977) and subsequent
studies of the MW Baade's Window by Rich and collaborators (e.g.\ Rich 1988). 
The high metallicities are properties typical of giant elliptical galaxies.  

The similarities of bulges and ellipticals strengthened the approach of
galaxian component analysis  whereby a galaxy can be analyzed into largely
independent structural units, foremost disks and  spheroids.  Spheroids would
encompass bulges and ellipticals, sharing similarities in structure, internal
dynamics and populations.  Such similarities would imply similar
formation histories and presumably old ages.  

\section{Surface brightness profiles}

There is now ample evidence that bulges do not follow the \rdeV\ law.  Kent et
al.\ (1991), using Spacelab IR telescope data, found that the vertical profile
of the MW bulge at 2.5 $\mu$m is well fit with an exponential law.  Andredakis
\& Sanders (1994) and de Jong (1996) showed that bulge-disk decompositions of
external galaxies led to more stable fits and smaller residuals when the
exponential law was used for bulges rather than the \rdeV\ law.  

\begin{figure}
\centering
\begin{minipage}[t]{0.45\textwidth}
  \raggedright
  \includegraphics[width=4.5cm]{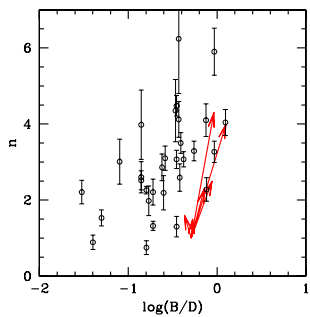}
\end{minipage}%
\begin{minipage}[t]{0.5\textwidth}
  \raggedleft
  \includegraphics[width=7cm]{exampleprofile.eps}
\end{minipage}\\[-10pt]
\begin{minipage}[t]{0.45\textwidth}
\raggedright
  \caption{\label{MBFig:NvsBD} {\it Points}: the distribution of 
    bulge S\'ersic index $n$ against the bulge-to-disk luminosity 
    ratio $B/D$ (APB95).  {\it Arrows}: growth 
    vectors for bulges derived from $N$-body simulations of satellite 
    accretion onto exponential bulges (Aguerri et al.\ 2001).  
    Arrows of increasing length correspond to 
    $M_{sat}/M_{bulge} = 1/6, 1/3, 1$, respectively.}
\end{minipage}%
\hspace{0.09\textwidth}
\begin{minipage}[t]{0.45\textwidth}
\raggedleft
  \caption{\label{MBFig:ExampleProfile} (a) $H$-band  HST+ground-based profile
  of NGC~5443 (Sb) fitted with PSF-convolved S\'ersic  bulge and exponential
  disk components.  (b) The same profile with a PSF-convolved point source
  (dashed line) added to the fitting function. Residuals in lower panels.  From
  Balcells et al.\ (2003a).}
\end{minipage}%
\end{figure}

A breakthrough in the characterization of bulge surface brightness profiles came
with the application of the S\'ersic (1968) \rSer\ model, 

\begin{equation}
I(r)=I_{e} \cdot \exp{\{-b_{n}\cdot [(r/r_{e})^{1/n}-1]}\},
\end{equation}

\noindent where $r_e$ is the effective radius, $I_e$ is the brightness at
$r_e$, $n$ is the {\sl shape index} and $b_{n}\approx 1.9992n-0.3271$  (Caon et
al.\ 1993).  APB95 showed that $K$-band bulge profiles are well fit with \rSer\
models, with a continuous distribution of shape indices, ranging from $n\approx
4-6$ for early types to $n\approx 0.5$ for late types
(Fig.~\ref{MBFig:NvsBD}).  Later fitting by G01, Khosroshahi et al.\ (2000a),
M\"ollenhoff \& Heidt (2001), and Trujillo et al.\ (2001) confirmed this
dependency.  The ability of S\'ersic's law to fit the range of structures of
bulges is comparable to its ability to fit elliptical galaxies (Caon et al.\
1993), suggesting a common range of structures for galaxian ellipsoids of a
wide range of luminosities.

Balcells et al.\ (2003a, hereafter B03) reanalyzed the sample
studied by APB95 using HST/NICMOS F160W images to extend the NIR profiles
inward to $\sim$0.1 arcsec.  At the increased resolution,
S\'ersic+exponential modeling yields strong central residuals: galaxy
centers are brighter than the inward extrapolation of the S\'ersic
model.  B03 show that for most cases (63\%) adding a single point
source (PSF-convolved) provides a good fit
(Fig.\ref{MBFig:ExampleProfile}).  In some cases, a central
exponential component, instead of a point source, improves the fit.
Such central components, with colors corresponding to
reddened stellar populations (Peletier et al.\ 1999), are largely
unresolved by the NICMOS Camera 2 (PSF = 0.19 arcsec FWHM).  Although the Nuker
model (Lauer et al.\ 1995) can provide good fits to the inner regions
of the profiles, out to a few arcsec, it cannot model the entire bulge
profile.  We address nuclear components in bulges in
\S\,\ref{MB:nuclei}.

Once nuclear components are included in the fitting function, the best-fit
Sersic models have considerably lower shape incides $n$ than obtained from
ground-based profiles.  For the B03 galaxies, the $n$ distribution has an upper
envelope that rises toward more luminous bulges, and a highest value of $n=3.0$
is found.  

Bulge structure shows strong links to internal dynamics.  Trujillo et al.\
(2001b) define a concentration index, \Cindex, as the ratio of luminosities
inside $r_{e}$ and inside $\alpha r_{e}$ ($\alpha < 1$).  For S\'ersic
profiles, \Cindex\ is a monotonic function of the shape index $n$.   \Cindex\
is useful because of its low sensitivity to the depth of the images.   
\Cindex\ is shown to correlate strongly with the central velocity dispersion
and with the mass of the Supermassive Black Hole inferred in galaxy centers
from kinematics measurements (Graham et al.\ 2001).  The correlation with
velocity dispersion opens an avenue  for measuring the Fundamental Plane of
spheroids using imaging data only (Graham 2002; see also Khosroshahi et al.\
2000b).  

\subsection{The scaling of bulge and disk parameters}
\label{MB:BulgeDiskScaling}

Studies of well-defined samples of disk galaxies have clarified that the
properties of bulges are intimately related to those of the disks.  While
common sense would suggest that earlier-type bulges are larger in relation to
their disks, it turns out that the ratio of bulge effective radius to disk
scale length is independent of Hubble type: $r_e/h \approx 0.1$ (de Jong 1996;
Courteau et al.\ 1996; Graham 2001; MacArthur et al.\ 2003).  Also, bulge
colors are very similar to the inner colors of their disks (Peletier \&
Balcells 1996).  These two relations are hard to establish if disks accreted
onto previously-formed old spheroids; rather, they suggest a coupled formation
of bulges and disks.

\section{Nuclear structures}
\label{MB:nuclei}

HST imaging at visible and NIR wavelengths has provided an important
increase in spatial resolution to study galaxy nuclei.  At
the distances of typical nearby samples ($cz \leq 3000$ km/s),
ground-based images (1'' FWHM seeing) resolve 150 pc, while WFPC2 (PSF
0.1'' FWHM) resolves nuclei down to $\sim$15 pc.  Bulges have been 
targeted in programs by Carollo and collaborators, and by Peletier and 
collaborators, in addition to many programs on ellipticals which also 
image S0 galaxies.  

At HST resolution, galaxy nuclei show a wealth of structure that contrasts with
the conception of bulges as smooth spheroids, including nuclear spirals, dust,
star formation and central sources often unresolved at HST resolution (Carollo
et al.\ 1998; Peletier et al.\ 1999; Ravindranath et al.\ 2001; Rest et al.\
2001; Seigar et al.\ 2002).  Because the nuclei of ellipticals show similar
levels of complexity (Phillips et al.\ 1995; Ravindranath et al.\ 2001), such
structures are often associated with the galaxy's bulge.  Their vertical sizes
are smaller than the disk vertical extent, hence we might associate them with
the bulge-spheroid or with the inner disk.

\begin{figure}
\centering
\begin{minipage}[t]{0.5\textwidth}
  \raggedright
  \includegraphics[width=6cm]{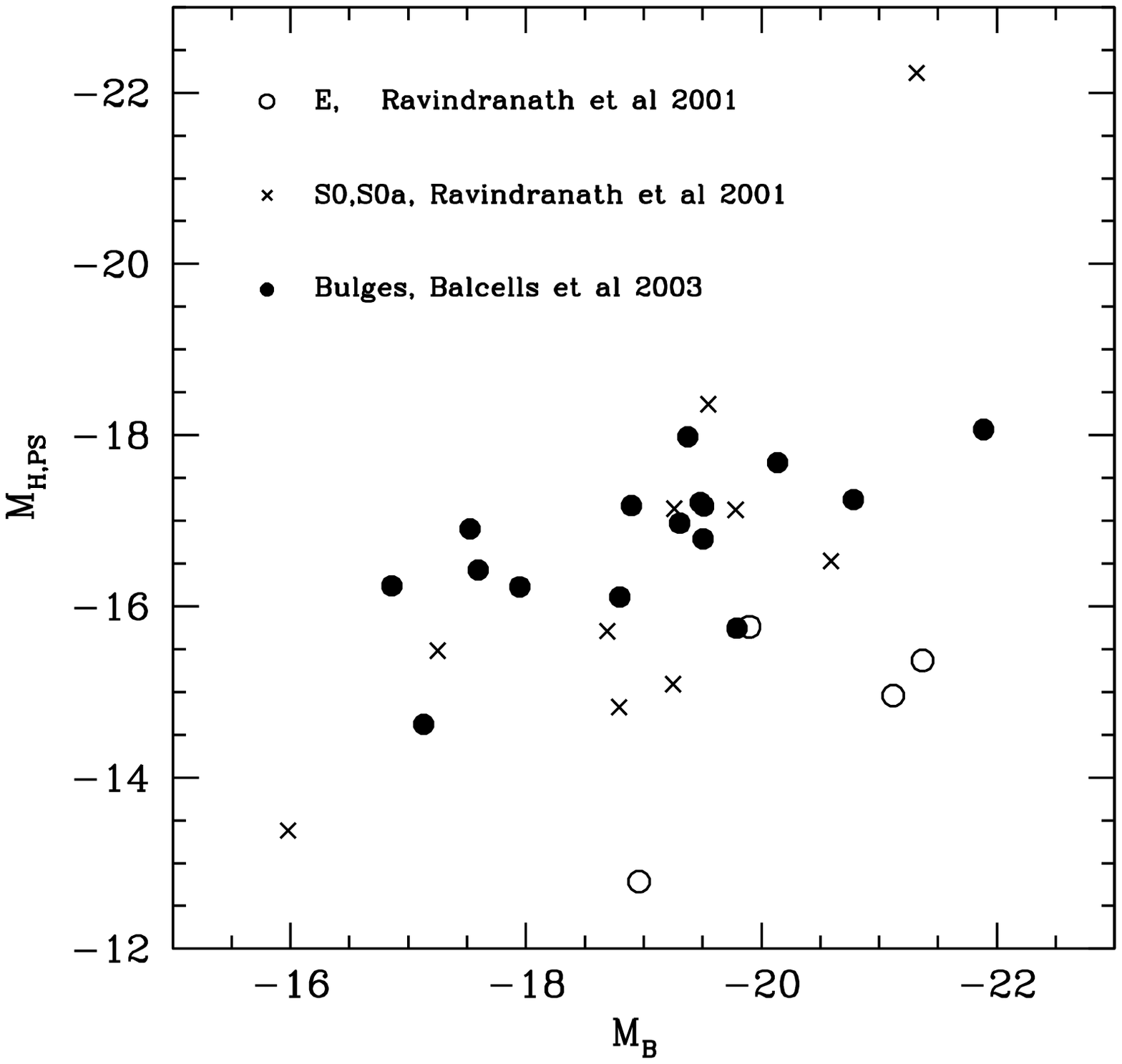}
\end{minipage}%
\begin{minipage}[t]{0.45\textwidth}
  \raggedleft
  \includegraphics[width=6cm]{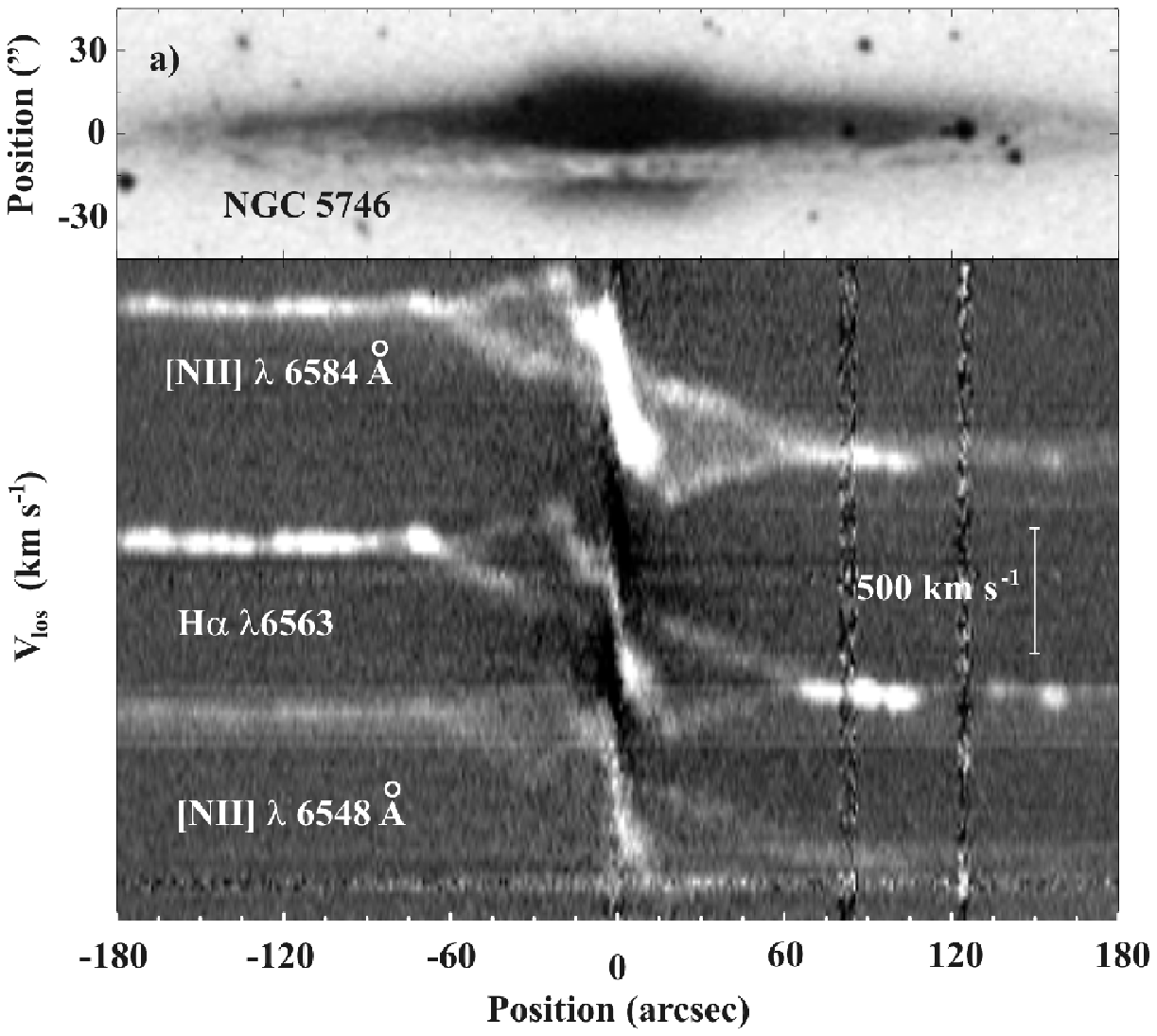}
\end{minipage}\\[-10pt]
\begin{minipage}[t]{0.45\textwidth}
\raggedright
  \caption{\label{MBFig:PSvsMB} $H$-band absolute magnitudes of  nuclear
  unresolved sources  against galaxy $B$-band absolute magnitudes, for samples
  of ellipticals  ({\it open circles}, Ravindranath et al.\ 2001), S0's ({\it
  crosses}, Ravindranath et al.\ 2001) bulges ({\it filled circles}, B03).  For
  the latter, bulge rather than total magnitudes are plotted. }
\end{minipage}%
\hspace{0.09\textwidth}
\begin{minipage}[t]{0.45\textwidth}
\raggedleft
  \caption{\label{MBFig:FigureOfEight} Position-velocity diagrams for the
  peanut-shaped galaxy NGC~5746,
  showing the figure-of-eight pattern given by gaps between orbit families in a
  bar (Bureau \& Freeman 1999).}
\end{minipage}%
\end{figure}

Unresolved sources have been detected in every study of galaxy centers using
HST imaging (Phillips et al.\ 1995; Faber et al.\ 1997; Carollo, Stiavelli \&
Mack 1998; Ravindranath et al.\ 2001; B03).  Figure~\ref{MBFig:PSvsMB} shows
point-source absolute magnitudes against galaxy absolute magnitude
from various sources.  Both the  frequency of detection and the inferred
luminosities depend on the  method applied to subtract the underlying bulge
light, so results  differ among the studies.  The frequency of detection ranges 
from 50\% to 85\%.  Point source luminosities are of order of  $10^{-3}$ of the
bulge luminosities, and, in B03, they scale as $L_{PS} \sim (L_{Bul})^{0.5}$.  

\section{Kinematics}
\label{MB:Kinematics}

Kinematic studies have provided strong evidence that not all bulges follow the
oblate-spheroid picture derived from $V/\sigma$ vs.\ $\epsilon$ studies
in the eighties (Kormendy \& Illingworth 1982).  

Kormendy (1993) summarizes the evidence  for disk-like dynamics in bulges. 
This includes flatness, low velocity dispersion and rapid rotation, as well as
spiral structure.   An example of the type of analysis involved is given in
Erwin et al.\ (2003) on the bulges of two barred galaxies (NGC~2787, NGC~3945).
The bulges show flattened light distributions with exponential surface
brightness profiles, and high rotation parameters $(V/\sigma)^\star \approx
1.5$, which hint that the light previously adscribed to their bulges comes
instead from luminous disks embedded in the bars.  Smaller, non-exponential,
rounder structures are found inside the inner disks.   Falc\'on-Barroso et al.\
(2003), working on a sample of 18 inclined disk galaxies, find that minor-axis
velocity dispersion gradients decrease with increasing bulge ellipticity.  Such
flat dispersion profiles in high-ellipticity bulges link smoothly with
dispersions in the disk, indicating a continuity of dynamical temperature from
bulge-dominated to disk-dominated regions of the galaxies.  This behavior is
not restricted to late types; it occurs in galaxies as early as S0 and with
high photometric B/D.  

Boxy or peanut-shaped bulges show a cylindrical velocity field, and very low
velocity dispersions except in a central peak (Shaw et al.\ 1993), linking
box-peanut bulges to thickened disks.  The connection of these bulges to
evolved disks was clearly established with the use of high-order kinematic
analysis.  Kuijken \& Merrifield (1995) and later Bureau \& Freeman (1999)
applied the Unresolved Gaussian Decomposition (UGD) technique to analyze gas
and stellar kinematics of edge-on, box-peanut bulges, showing the
figure-of-eight pattern in the position-velocity diagrams expected from X1 and
X2 orbits in bars (Fig.~\ref{MBFig:FigureOfEight}).  That such a feature does
not appear in bulges with elliptical isophotes strongly links box-peanut bulges
to bars.  

The above studies reveal that some photometric bulges are really
disks with brightness in excess of the inward extrapolation of the outer
exponential disk.   The frequency of kinematically disky bulges can be
established by studying well-defined samples with the  upcoming generation of
3D spectrographs, e.g. SAURON (de Zeeuw et al.\ 2002).  

Bulges rotate around the same axis as the disk does, i.e.\ they do not show minor
axis rotation (Falc\'on-Barroso et al.\ 2003), with a few notable exceptions.  As an
example, the bulge of NGC~4698 rotates about an axis perpendicular to the disk
rotation axis (Bertola et al.\ 1999).  These objects indicate that accretion or
merging has been important for some bulges.  The misaligned fraction may be
related to the frequency of accretion events.   However, while monolithic
collapse is unlikely to lead to disk-bulge kinematic misalignment, mergers do
not necessarily generate kinematic misalignment.  Hence, the fraction of
misaligned bulges is a lower limit to the fraction of bulges with a rich
accretion history.  

\subsection{New kinematic spectral analysis techniques}

Falc\'on-Barroso et al.\ (2003) have succesfully used single-burst stellar
population (SSP) models as templates for kinematic spectral analysis.  This has
become possible thanks to the development of new SSP NIR spectral energy
distributions (SED) at 1.5\AA\ resolution (Vazdekis et al.\ 2003; see also
Vazdekis 1999).  The method consists of measuring the spectrograph's broadening
function in calibration arc exposures, and convolving the synthetic spectral
templates to match that broadening.  Determining galaxy kinematics can often be done
better with SSP templates than with stellar templates, probably as a
result of reduced template mismatch.  The method allows a reduction in
telescope time requirements, and has strong potential for high-redshift galaxy
kinematics, where target galaxy and stellar templates would be observed in
different wavelength ranges.  

\section{Fundamental Plane, and bulge population ages}
\label{MB:FP}

The distribution of a galaxy sample on the Fundamental Plane (FP) defined by
old ellipticals can in principle provide three types of diagnostics. 
The FP is usually expressed as 

\begin{equation}
	\label{MBeq:VT}
    \log(r_{e}) = \alpha \log(\sigma_0) + \beta <\mu>_e + \gamma
\end{equation}

\noindent while, assuming Virial equilibrium, one has 

\begin{equation}
	\label{MBeq:VTFP}
    \log(r_{e}) = 2\,\log(\sigma_0) + 0.4\,<\mu>_e + \log(C_{r}C_{v}) -
    \log(\frac{M}{L}) +  \gamma'
\end{equation}

\noindent (Djorgovski, de Carvalho \& Han 1988; Bender et al.\ 1992; Graham \&
Colless 1995), where variations of the structural and kinematic parameters
$C_{r}$ and $C_{v}$ reflect non-homology.  First, the thinness of the
distribution provides a measure of the sample homogeneity. Second,  if one
believes that all the objects in the sample are structurally and dynamically
homologous, $M/L$ trends with luminosity may be inferred (leading to $M/L \sim
L^{0.25}$, Faber et al.\ 1987).  This provides population age trends with
luminosity.   Finally, one may investigate whether broken homology can
reconcile the FP with the VT prediction without recourse to $M/L$ variations
(Hjorth \& Madsen 1995; Capelato et al.\ 1995).   

Key difficulties in establishing the FP of bulges are the disk contamination to
the surface brightness and the velocity dispersion, and dust extinction. 
Falc\'on-Barroso et al.\ (2002) address these problems by measuring structural
parameters from $K$-band images, and using colors unaffected by dust to convert
those to the standard $B$-band.  These authors find a small but significant
offset of early-to-intermediate type bulges with respect to the Coma
ellipticals of J\"orgensen et al.\ (1996).  Their result rests on an exquisite
attention to details such as aperture corrections for the velocity
dispersions.  Similar offsets had been  found by Bender et al.\ (1993) using
$B$-band imaging data.  

Can the FP offset of bulges w.r.t.\ Coma ellipticals be due to  non-homology
rather than to population differences?  Gonz\'alez (2003) shows  that, for a
range of ellipsoids produced by collisionless mergers of spirals, $C_{r}$ and
$C_{v}$ (eqn.~\ref{MBeq:VTFP}) do vary from object to object, but the product
$(C_{r}C_{v})$ remains constant among merger remnants.  The structure and
internal dynamics conspire to bring non-homologous ellipsoids onto the same FP
(see also Trujillo, Graham, \& Caon 2001), implying that merger-driven
non-homology may not contribute to offsets or thickening of the FP.  In terms
of kinematic non-homology only, Falc\'on et al.\ (2002) find that the non
inclusion of a rotational support term in the FP can explain the observed
offset of bulges w.r.t. ellipticals.  

The FP results are generally interpreted as evidence that bulges are, on the
whole, slightly younger than cluster ellipticals.  By how much?  When assuming
homology and ignoring rotation, Falc\'on et al.\ (2002) derive bulge ages of
around 10 Gyr if ellipticals are 12.5 Gyr old.   This result is consistent with
those obtained from colors and from line strengths.  Optical colors suffer the
well-known age-metallicity degeneracy.  In bulges, dust in the disk only
aggravates the problem.  The degeneracy can however be partially broken by the
combined use of optical and NIR colors, e.g. $U-R$ vs.\ $R-K$ or $B-I$ vs.\
$I-H$ (Bothun \& Gregg 1990; Peletier \& Balcells 1996; Peletier et al.\ 1999).
$U-R$ is sensitive to age and dust, while $R-K$ is sensitive to dust and
metallicity.  Fitting single-stellar population (SSP) models to colors on
dust-free lines of sight, Peletier et al.\ (1999) infer bulges ages above
$10^{10}$ Gyr, at most about 2 Gyr younger than the ages of Coma ellipticals. 
Most importantly, no clearly young bulges are found in a sample of early- to
intermediate-type bulges (Sbc galaxies do show slightly younger ages). 
Line-strength indices yield similar results; in an ongoing program of
minor-axis spectroscopy of bulges, using H$\gamma$ to disentangle age from
metallicity, Goudfrooij, Gorgas, \& Jablonka (1999) also find bulge population
ages slightly younger than cluster ellipticals, and with a similar age span as
field ellipticals.

Dating bulge populations to answer who formed first, the bulge or the disk, is
a subtle problem.  Peletier \& Balcells (1996) infer that bulges are older than
the inner regions of the disks by at most $\Delta\log\,t = 0.11$ ($\Delta\,t =
3$ Gyr for $t_{disk}=10$Gyr), assuming equal metallicities, and by much less if
bulge metallicities are higher.  One may argue that SSP models applied to local
galaxy samples are too simplistic to resolve the small age differences between
bulges and diks, but the results may not be too far off.   The ages given
above would put the bulk of the star formation in bulges and disks at redshifts
$z \geq 1.5$, a result that is consistent with the onset of structurally mature
disk galaxies at $z\sim 1$ (Lilly et al.\ 1998).   A powerful avenue to improve
these results will be to study epochs closer in time to the epoch when the
bulge-disk strucures assembled.  This demands studying galaxy samples
approaching $z=1$, something now feasible with recourse to HST optical and NIR
imaging (e.g. Ellis et al.\ 2001).

\section{Formation models}
\label{MB:Formation}

\subsection{Accretion}
\label{MB:Accretion}

In a {\sl soft} version of the merger origin proposed by CDM-based galaxy
formation models, bulges may form/grow by accreting smaller structural units. 
Evidence for accretion comes from, e.g., orthogonal bulge kinematics
(\S\,\ref{MB:Kinematics}).  Orbital decay
by dynamical friction will indeed feed the galaxy's central bulge.  The
question is whether the numbers add up: i.e.\ did satellites exist with the
right numbers and the right masses for dynamical friction orbital decay to
build up a bulge mass in a Hubble time?.  Fu et al.\ (2003) investigate orbital
decay of Super Star Clusters (SSC) to explain the formation of small, late-type
bulges with cuspy cores.  At the other end of the B/D spectrum, accretion
formation models for large bulges were studied by Pfenniger
(1993).  He modeled the accretion of several satellites onto a disk-halo
galaxy.  The satellites end up at the core, while the disk thickens to a
Sombrero-like shape.  His model, with rigid satellites which do not lose mass,
is probably simplistic; nevertheless, it highlights that bulges grown by
accretion are not composed of stars from the accreted galaxies only, and that
other mechanisms besides bar buckling instabilities may pump disk material
up into a bulge-like distribution.

Aguerri, Balcells, \& Peletier (2001) investigate the accretion hypothesis by
checking whether bulges grown by accreting satellites follow the trend of
S\'ersic index $n$ with $B/D$ observed by APB95.  They study whether the
accretion of satellites onto model galaxies harboring exponential bulges
provides not only additional mass to the bulge but also a structural change
toward higher-$n$ S\'ersic profiles.  The results of the $N$-body models are
shown in Figure~\ref{MBFig:NvsBD}, with arrows tracing growth vectors for
bulges in the $n$--$B/D$ plane, together with the points from APB95.  It is
promising that the trends are the same, although growth of $n$ appears steeper
in the models.  As it turns out, exponential profiles are very fragile to
merging.  Hence, it is unlikely that bulges with exponential profiles grew
significantly through collisionless mergers.  The models have several
limitations.  The disks thicken during  satellite accretion, and rebuilding of
a thin disk needs to be  postulated (the thick disk left over after the
accretion may be related to the thick disks of early-type disk galaxies).  It
is not clear how bulge growth via mergers plus rebuilding  of the thin disk can
explain the observed scaling of bulge and disk  parameters
(\S\,\ref{MB:BulgeDiskScaling}).  Accretion of satellites could explain the
class of distorted, thick boxy bulges studied by L\"utticke  et al.\ (2000).  

\subsection{Coupled bulge-disk formation models}
\label{MB:vdBosch}

Van den Bosch (1998, 2000) included bulges in the Fall \& Efstathiou (1980)
theory of disk formation.  Gas cooling inside a virialized halo is highly
unstable and forms a central bulge component rather than a disk.  When the
bulge has grown sufficiently, its central potential allows subsequent gas to
settle into an ordered disk.  The stability condition determines a higher
bulge-to-disk ratio at increasing formation redshift.  This model broadly
accounts for the ages of bulges and disks, and allows for the formation of a
central bulging component without recourse to merging; the bulge stars did not
dominantly form in disks as they do in the merger hypothesis.  

\section{Conclusions}
\label{MB:Conclusions}

The ability of the S\'ersic model to describe the  shapes of the surface
brightness profiles of both bulges and ellipticals suggests that bulges and
ellipticals belong to a single family of ellipsoidal galaxian components.  The
proximity of the FP of both bulges and ellipticals also emphasizes their
similar structure.  Merger models account for some of these properties.   On
the other hand, both the detailed structure and kinematics of some bulges of
all Hubble types display properties more typical of thickened disks than of
oblate spheroids.   Kinematic studies of large,
well-defined samples will be useful for determining the prevalence of disky
"bulges" across the Hubble sequence and the bulge luminosity sequence.  Once we
know how to distinguish disky from classical spheroidal bulges, a new term
might be desirable to refer to disky bulges.

Structural parameters and colors of bulges and disks are coupled, arguing
against a clean separation of the epochs and formation mechanisms for bulges
and disks.  Population diagnostics show that populations in early-type bulges
are old, with ages above $10^{10}$ Gyr, approaching those of cluster
ellipticals.  These two results place the assembly of disk-bulge galaxies at
redshifts $z\geq 1.5$.  

\smallskip

\noindent {\bf Acknowledgements:} I would like to thank R. Peletier, A. Graham and
P. Erwin for discussions and for comments on the manuscript.


\end{document}